\documentclass{tlp}

\usepackage{amssymb}
\usepackage{url}

\newcommand{\sig}{\mathcal{L}}
\newcommand{\lif}{\leftarrow}
\newcommand{\liff}{\leftrightarrow}
\newcommand{\lthen}{\rightarrow}
\newcommand{\lbox}{{\cal K}}
\newcommand{\ldiamond}{{\cal B}}
\newcommand{\prove}{\vdash}
\newcommand{\cprove}{\Vdash}

\newcommand{\set}[1]{\left\{#1\right\}}
\newcommand{\tq}{\;|\;}
\newcommand{\name}[1]{\mathrm{#1}}
\newcommand{\Gi}[1][i]{{\name{G}_{#1}}}
\newcommand{\equivin}[1]{\equiv_\name{#1}}
\newcommand{\provein}[1]{\prove_\name{#1}}
\newcommand{\cprovein}[1]{\cprove_\name{#1}}
\newcommand{\setcomp}[1]{\widetilde{#1}}

\newcommand{\RedOne}[1]{\name{Redu1}(#1)}
\newcommand{\RedTwo}[1]{\name{Redu2}(#1)}
\newcommand{\GenDisj}[1]{\name{GenDis}(#1)}
\newcommand{\AugFree}[1]{\name{AugFree}(#1)}
\newcommand{\FreeGen}[1]{\name{FreeGen}(#1)}

\newenvironment{lpblock}[1][l]{\begin{quote}$\begin{array}{#1}}{\end{array}$\end{quote}}

\newtheorem{theorem}{Theorem}[section]
\newtheorem{proposition}[theorem]{Proposition}
\newtheorem{lemma}[theorem]{Lemma}
\newtheorem{corollary}[theorem]{Corollary}
\newtheorem{remark}[theorem]{Remark}

\newtheorem{definition}{Definition}[section]
\newtheorem{example}[definition]{Example}

\title[Applications of Intuitionistic Logic in Answer Set Programming]%
  {Applications of Intuitionistic Logic in\\ Answer Set Programming}
  
\author[M. Osorio, J. A. Navarro and J. Arrazola]{MAURICIO OSORIO,
        JUAN A. NAVARRO AND JOS\'E ARRAZOLA \\
        Universidad de las Am\'ericas, CENTIA,\\
        Sta. Catarina M\'artir, Cholula, Puebla\\
        72820 M\'exico \\
        \email{josorio@mail.udlap.mx}
       }
       
\begin{document}
\maketitle

\begin{abstract}
We present some applications of intermediate logics in the field of Answer 
Set Programming (ASP). A brief, but comprehensive introduction to the answer set 
semantics, intuitionistic and other intermediate logics is given. Some 
equivalence notions and their applications are discussed. Some results on 
intermediate logics are shown, and applied later to prove properties of answer 
sets. A characterization of answer sets for logic programs with nested 
expressions is provided in terms of intuitionistic provability, generalizing a 
recent result given by Pearce.

It is known that the answer set semantics for 
logic programs with nested expressions may select non-minimal models. Minimal 
models can be very important in some applications, therefore we studied them; in 
particular we obtain a characterization, in terms of intuitionistic logic, of answer
sets which are also minimal models. We show that the logic $\Gi[3]$ 
cha\-rac\-te\-ri\-zes the notion of strong equivalence between programs under 
the semantic induced by these models. Finally we discuss possible applications and
consequences of our results. They clearly state interesting links between ASP and
intermediate logics, which might bring research in these two areas together.
\end{abstract}


\section{Introduction}
Answer Set Programming (ASP), Stable Logic Programming or 
A-Prolog, is the realization of much theoretical work on Non-monotonic Reasoning 
and AI applications of Logic Programming (LP) in the last 15 years. The main 
syntactic restriction needed in this paradigm is to eliminate function symbols 
from the language. This is because using infinite domains the
answer sets are no longer necessarily recursively enumerable \cite{MarRem01}. 
The two most well known systems that compute answer sets are
\texttt{dlv}\footnote{\url{http://www.dbai.tuwien.ac.at/proj/dlv}} and 
\texttt{smodels}\footnote{\url{http://saturn.hut.fi/pub/smodels}}.

Our work is intended to provide an alternative view of the theory of answer set 
programming through different tools and relations with intuitionistic and other 
intermediate logics. We provide a characterization of answer sets by 
intuitionistic logic as follows: 
\begin{quote}\it
``A formula is entailed by a logic program in the answer set semantics if and 
only if it can be proved in every intuitionistically complete and consistent 
extension of the program formed by adding only negated literals.''
\end{quote}
This is a generalization of a recent result given by Pearce where he considered 
disjunctive programs only. In our approach we consider the class of augmented 
programs, which allow nested formulas in the head and the body of
clauses. \citeN{ErdLif01} provided some evidence on how
augmented programs can be used to represent and solve real life problems.

Our result provides foundations of defining the notion of non-monotonic in\-fe\-rence of
any propositional theory (using the standard connectives
$\set{\lnot, \land, \lor, \lthen}$) in terms of a monotonic logic (namely intuitionistic
logic). We propose the fo\-llow\-ing interpretation: We understand the \emph{knowledge}, of
a given theory $T$, as all the formulas $F$ such that $F$ is derived from $T$ using
intuitionistic logic. This makes sense since in intuitionistic logic, according to
\citeN{Bro81}, $A$ can be interpreted as ``I know  $A$''. We will also identify a
set of \emph{beliefs} for the theory $T$. We will say it is safe to believe a formula $F$
if and only if $F$ belongs to every intuitionistically complete and consistent extension of
$T$ by adding only negated literals.

Take for instance: $\lnot a \lthen b$. The agent knows $\lnot a \lthen b$, 
$\lnot b \lthen \lnot\lnot a$ and so on. The agent, however, does not know neither
$a$ nor $b$. Nevertheless, one believes more than one knows. But a cautious agent must
have his/her beliefs consistent to his/her knowledge. This agent will try to assume 
negated literals in order to infer more information. Thus, in our example, our 
agent can believe $\lnot a$, since this assumption is consistent, in order to conclude
$b$. At this point the agent can decide, for any formula constructed from $a$ and $b$,
either if it is true or false. The theory is now complete.

The agent could also try to assume $\lnot b$ in order to conclude $\lnot\lnot a$, but
he/she would not be able to intuitionistically prove $a$ and the theory can not be
completed. Thus it was not safe to believe $\lnot b$. It also makes sense that a 
cautious agent could try to believe $\lnot\lnot a$ rather than to believe 
$a$ (recall that $a$ is not equivalent to $\lnot\lnot a$ in intuitionistic 
logic). Our results agree with the position of \citeN{Kow01}, namely
``that Logic and LP need to be put into place: Logic within the thinking
component of the observation-thought-action cycle of a single agent, and LP
within the belief component of thought''.

One important issue to know is when two programs are ``equivalent'' with respect 
to the answer set semantics. We consider a definition for ``equivalence'' that 
is given in \citeN{LiPeVa01}. We say that $P_1$ and $P_2$ are strongly equivalent 
if for every program $P$, $P_1 \cup P$ and  $P_2 \cup P$ have the same answer 
sets. If two programs are strongly equivalent, we know that we can replace one 
by the other in any larger program without changing the declarative semantics. 
This is an important concept for software engineering. It has been shown that 
the logic of Here-and-There (HT) or $\Gi[3]$ characterizes the class of strongly 
equivalent augmented programs under this definition \cite{LiPeVa01}.

If we want to use a program $P_2$ instead of another one $P_1$, it will be 
perfect if both programs have the same answer sets. This condition is, however,
sometimes too much to expect. It will suffice if we can identify,
through a simple relation, the answer sets of the first program knowing
those of the second. A conservative extension \cite{OsNaAr:lopstr} is one form
of this weaker type of equivalence.

In order to define a ``strong'' version of this notion of equivalence we find 
useful to split the signature of programs into some user atoms and reserved 
atoms. The idea is that users are allowed to write programs using only the user 
atoms, while reserved atoms are used for internal program transformations. Given 
a user program $P_1$ and an internal program $P_2$, we say that $P_2$ is a 
strong conservative extension of $P_1$ if for every user program $P$, it holds 
that $P_2 \cup P $ is a conservative extension of $P_1 \cup P$. We show then 
that for every augmented program $P$ there is a disjunctive program $P'$ such 
that $P'$ is a strong conservative extension of $P$. We also illustrate how to 
compute such a program $P'$.

Minimal models are of general interest for several theoretical and practical 
reasons \cite{BeNeNg93,GePrPr89,Lib99,LobSub92,MinPer85}. We 
therefore devote a section to study them in the context of answer sets. We first
provide a characterization, in terms of intuitionostic logic, of answer sets that
are also minimal models. And we show that two programs are strongly
equivalent, with respect to the induced semantic, if and only if they are
equivalent in the 3-valued logic $\Gi[3]$.

In this paper we restrict our attention to finite propositional theories; the 
semantics can be extended to theories with variables by grounding. Function
symbols are, however, not allowed to ensure the ground program to be finite.
This is a standard procedure in ASP. We assume that the reader has some basic
background in logic and Answer Set Programming.

Our paper is structured as follows: In Section~\ref{Back} we present the general 
syntax of clauses and define several types of programs. We also provide the 
definition of answer sets for augmented logic programs as well as some
background on logic. In Section~\ref{sect-eq} we present our notions of
equivalence and provide some useful transformations to simplify the structure
of programs. In Section~\ref{sect-appl} we present our main result, the
characterization of answer sets in terms of intuitionistic logic. In
Section~\ref{sect-minimal} we study the class of answer sets that are minimal models.
In Section~\ref{Future} we discuss several interesting consequences of the
proposed approach and our main result. In Section~\ref{conclusion}, we present
some conclusions and ideas for future work. Finally as an appendix, in
Section~\ref{appendix-proofs}, we present the proofs of our results.

\section{Background}\label{Back}

In this section we review some basic concepts and definitions that will be used 
along this paper. We introduce first the syntax of formulas and programs based 
on the language of propositional logic. We also describe some common classes of 
logic programs and give the definition of answer sets. Finally we make some 
comments on intermediate logics that will be used in later sections to study
the notions of answer sets and non monotonic reasoning.

\subsection{Propositional Logic}
We use the language of propositional logic in order to describe rules within 
logic programs. Formally we consider a language built from an alphabet
consisting of \emph{atoms}: $p_0, p_1, \dots$;
\emph{connectives}: $\land, \lor, \lif, \bot$; and 
\emph{auxiliary symbols}: `$($', `$)$', `$.$'.

Where $\land, \lor, \lif$ are 2-place connectives and $\bot$ is a 0-place 
connective.
Formulas are defined as usual. The formula $\top$ is introduced as
an abbreviation of $\bot \lif \bot$, $\neg F$ as an abbreviation of
$\bot \lif F$, and $F \liff G$ as an abbreviation of $(G \lif F) 
\land (F \lif G)$. The formula $F \lthen G$ is  another way of writing the 
formula $G \lif F$, we use the second form because of tradition in the context 
of logic programming.

A signature $\sig$ is a finite set of atoms. If $F$ is a formula then the 
\emph{signature} of $F$, denoted as $\sig_F$, is the set of atoms that occur in 
$F$. A \emph{literal} is either an atom $a$ (a positive literal) or a negated 
atom $\neg a$ (a negative literal). A \emph{theory} is just a set of formulas.

\subsection{Logic Programs}
A \emph{logic program} is a finite set of formulas. The syntax of formulas 
within logic programs has been usually restricted to clauses with very simple 
structure. A \emph{clause} is, in general, a formula of the form $H \lif B$ 
where $H$ and $B$ are known as the \emph{head} and \emph{body} of the clause 
respectively. Two particular cases of clauses are \emph{facts}, of the form $H 
\lif \top$, and \emph{constraints}, $\bot \lif B$. Facts and constraints are 
sometimes written as $H$ and $\lif B$ respectively.

We introduce several kinds of clauses commonly found in literature.
A \emph{free clause} is built from a disjunction of literals in the head and
a conjunction of literals in the body. Such a clause has the form
$$h_1 \lor \cdots \lor h_n \lif b_1 \land \cdots \land b_m.$$
where each $h_i$ and $b_j$ is a literal. Either the head or the body of a free 
clause could be empty to denote a constraint or a fact.
A \emph{general clause} is a free clause that does not allow negation in the 
head, all literals in the head of the clause should be positive atoms.
Finally, a \emph{disjunctive clause} is a general clause with a non-empty
head, i.e.\ it is not a constraint.

A \emph{nested formula} is a formula built from the connectives $\land$, $\lor$
and $\lnot$ arbitrarily nested. An \emph{augmented clause} is a less restricted
form of clause where both $H$ and $B$ can be nested formulas.
Note, however, that embedded implications are not allowed in augmented
clauses. The formula $a \lif (b \lthen c)$ is not, for instance,
an augmented clause. The following are examples of clauses just defined
\begin{lpblock}[l@{\qquad}l]
a \lor b \lif c \land d \land \lnot e. & \mbox{disjunctive, general, free, augmented} \\
\bot \lif p \land q.                   & \mbox{general, free, augmented (constraint)} \\
a \lor \lnot b \lif p \land \lnot q.   & \mbox{free, augmented} \\
a \lor \lnot a.                        & \mbox{free, augmented (fact)} \\
\lnot (p \land \lnot q) \lif a \lor (\lnot b \land c). & \mbox{augmented}
\end{lpblock}

We also say that a logic program is free if it contains only free clauses.
Similarly, disjunctive and augmented programs are introduced. We would
also use the term \emph{logic program} alone to denote a set of arbitrary
propositional formulas with no restrictions at all.

\subsection{Answer sets}

We present now the definition of answer sets for augmented programs. This 
material is taken from \citeN{LiTaTu99} with minor modifications since they 
consider a broader syntax of formulas. They consider two kinds of negation: 
default and classical. Our negation $\lnot$ corresponds to their default 
negation $not$. Classical negation is not considered since it is easy to 
simulate it using a proper renaming of atoms. They also include an if-then-else 
constructor, but it is only an abbreviation of another formula. Hence, it is fair 
to say that their programs extend our augmented programs only by allowing the 
use of `classical' negation. 

Atoms, as well as the connectives $\bot$ and $\top$, are called
\emph{elementary formulas}. Formulas built from $\land$ and $\lor$ over
elementary formulas are called \emph{basic}. Similarly basic clauses and
programs are constructed from basic formulas. The definition of answer sets
is given first for basic programs, without default negation, and is extended
later to the class of augmented programs \cite{LiTaTu99}.

\begin{definition}\cite{LiTaTu99}
We define when a set of atoms $X$ \emph{satisfies} a basic formula $F$, denoted 
$X \models F$, recursively as follows:
\begin{description}
\item for elementary $F$, $X \models F$ if $F \in X$ or $F = \top$.
\item $X \models F \land G$ if $X \models F$ and $X \models G$.
\item $X \models F \lor G$ if $X \models F$ or $X \models G$.
\end{description}
\end{definition}

Note that the previous definition does not contain the case of implication,
since the syntax of augmented formulas does not allow to embed them as a
subformula. Only one implication is allowed in each clause, and this is taken
into account in the next definition.

\begin{definition}
\cite{LiTaTu99}
Let $P$ be a basic program. A set of atoms $X$ is \emph{closed} under $P$ if,
for every clause $H \lif B \in P$, $X \models H$ whenever $X \models B$.
\end{definition}

\begin{definition}\cite{LiTaTu99}
Let $X$ be a set of atoms and $P$ be a basic program. $X$ is an \emph{answer 
set} of $P$ if $X$ is minimal among the sets of atoms closed under $P$.
\end{definition}

\begin{definition}\cite{LiTaTu99}
The \emph{reduct} of an augmented formula or program, relative to a set of
atoms $X$, is defined recursively as follows:
\begin{description}
\item for elementary F, $F^X = F$.
\item $(F \land G)^X = F^X \land G^X$.
\item $(F \lor G)^X = F^X \lor G^X$.
\item $(\lnot F)^X = \bot$ if $X \models F^X$ and
$(\lnot F)^X = \top$ otherwise.
\item $(H \lif B)^X = H^X \lif B^X$.
\item $P^X = \set{(H \lif B)^X  \tq H \lif B \in  P}$.
\end{description}
\end{definition}

Observe that the reduct of an agumented program, obtained as in previous
definition, is a basic program. Using this reduct operator we are able to
extend the definition of answer sets to the class of augmented programs.

\begin{definition}[Answer Sets]\cite{LiTaTu99}\label{answersetdef}
Let $P$ be an augmented program and $X$ be a set of atoms. $X$ is an 
\emph{answer set} of $P$ if it is an answer set of the reduct $P^X$.
\end{definition}

\begin{example}\label{ansetsample}
Consider the  following program $P$:
\begin{lpblock}
a \lif \neg\neg a. \\
\neg b \lif c \lor b.
\end{lpblock}
If we take $X = \set{a}$ then the reduct is $P^X$:
\begin{lpblock}
a \lif \top. \\
\top \lif c \lor b.
\end{lpblock}
Here it is easy to verify that $\set{a}$ is closed under this reduct and,
since the empty set $\emptyset$ is not, it is the minimal set with this property.
Then it follows that $\set{a}$ is an answer set of $P$. However note that the
empty set $\emptyset$ is also an answer set of $P$, since it produces a different
reduct and is closed under it.
\end{example}

\subsection{Intermediate Logics}

The main goal of the research presented in this paper is to study the current 
definition of answer sets in terms of mathematical logic. We present an 
extremely simple, logical characterization of answer sets applicable to
augmented programs, based on a well-known alternative to classical logic,
namely intuitionistic logic. Several interesting consequences of our
approach are discussed in more detail in Section~\ref{Future}.

We briefly describe in the following lines multivalued and intuitionistic
logics. Interesting relations between these logics and the answer set
semantics are studied in later sections. Some notation, definitions and simple
results are given at the end of this section.

\subsubsection{G\"odel Multivalued Logics.}
These logics are defined generalizing the idea of truth tables and evaluation 
functions of classical logic. G\"odel defined the multivalued logics $\Gi$, with 
values in $\set{0, 1, \dots, i-1}$, with the following evaluation function $I$:
\begin{itemize}
\item $I(B \lif A) = i-1$ if $I(A) \leq I(B)$ and $I(B)$ otherwise.
\item $I(A \lor B) = \max(I(A), I(B))$.
\item $I(A \land B) = \min(I(A), I(B))$.
\item $I(\bot) = 0$.
\end{itemize}
An interpretation is a function $I\colon \sig \to \set{0, 1, \dots, i-1}$ that 
assigns a truth value to each atom in the language. The interpretation of an
arbitrary formula is obtained by propagating the evaluation of each connective as
defined above. Recall that $\lnot$ and $\top$ were introduced as abbreviations
of other connectives. An interpretation is said to be \emph{definite} if it
assigns only values $0$ or $i-1$, and \emph{indefinite} if some intermediate
value is assigned to an atom.

For a given interpretation $I$ and a formula $F$ we say that $I$ is a 
\emph{model} of $F$ if $I(F) = i-1$. Similarly $I$ is a \emph{model} of a 
program $P$ if it is a model of each formula contained in $P$. If $F$ is
modeled by every possible interpretation we say that $F$ is a \emph{tautology}.
Notice that $\Gi[2]$ coincides with classical logic 
$\name{C}$. The 3-valued logic $\Gi[3]$ is particularly useful for some
of our results.

\subsubsection{Intuitionistic Logic.}
This is an important logic, which has been an area of great interest during the 
last years. It is based on the concept of \emph{proof} or 
\emph{know\-led\-ge}, rather than \emph{truth} in classical logic, 
to explain the meaning and use of logical connectives.

Intuitionistic logic, denoted $\name{I}$, can be defined in terms of Hilbert 
type proof systems of axioms and inference rules.
Equivalent definitions can be given in terms of natural 
deduction systems and Kripke models \cite{TroDal88,Dal80}. 
Surprisingly no definition using a truth table scheme is possible.
Provable formulas are called \emph{theorems}.
G\"{o}del observed that there are infinitely many logics located between 
intuitionistic and classical logic \cite{ZaWoCh01}. In particular it has been
shown that
$$\name{I} \subset \cdots \subset \Gi[i+1] \subset \Gi \subset \cdots \subset 
\Gi[3] \subset \Gi[2] = \name{C}$$
where $\subset$ denotes proper inclusion of the set of provable formulas on 
each logic.
We use the term \emph{intermediate logic} to denote all logics, sets of classical
tautologies closed under modus ponens and propositional substitution, that contains
all the intuitionistic theorems. We say that a logic is a \emph{proper intermediate
logic} if it is an intermediate logic and is not the classical one. Observe that
the multivalued logics $\Gi$ are intermediate logics.

\subsubsection{Notation and General Definitions}
We use the standard notation $\provein{X} F$ to denote that $F$ is provable (a 
tautology, a theorem) in logic $\name{X}$. If $T$ is a theory we 
understand the symbol $T \provein{X} F$ to mean that $\provein{X} F \lif 
(F_1 \land \cdots \land F_n)$ for some formulas $F_i$ contained in 
$T$. This is not the usual definition given in literature, but can be shown 
to be equivalent because of results like the \emph{Deduction Theorem}.
Similarly, if $U$ is a theory, we use the symbol $T \provein{X} U$ to denote
$T \provein{X} F$ for every $F \in U$.

A theory $T$ is said to be consistent, with respect to logic $\name{X}$, if it 
is not the case that $T \provein{X} \bot$. Also, a theory $T$ is said to be
(literal) \emph{complete} if, for every atom $a \in \sig_T$, we have either
$T \provein{X} a$ or $T \provein{X} \lnot a$. We say that a program is
\emph{incomplete} if it is not complete.

We use the notation $T \cprovein{X} U$ to stand for the phrase: $T$ is
consistent and $T \provein{X} U$. Finally we say that two theories $T_1$ and
$T_2$ are equivalent under logic $\name{X}$, denoted by $T_1 \equivin{X} T_2$,
if it is the case that $T_1 \provein{X} T_2$ and $T_2 \provein{X} T_1$.

\section{Equivalence Notions}\label{sect-eq}

Given two programs we find useful to define several forms of equivalence
relations. The most natural equivalence notion that can be defined in terms
of the answer set semantics is that two programs are \emph{equivalent} if they
have exactly the same answer sets. However, this notion of equivalence is, 
sometimes too weak since it does not satisfy certain properties we would
expect from an equivalence relation. Some other equivalence notions with
richer properties need to be defined.

\subsection{Strong Equivalence}

Observe that, for instance, replacing equivalent pieces of programs in a larger program
does not always ensure that the original and the transformed program are equivalent.
The notion of \emph{strong equivalence} is defined looking for this kind of properties.

\begin{definition}\label{strongequiv}\cite{LiPeVa01}
Two programs $P_1$ and $P_2$ are \emph{strongly equivalent} if $P_1 \cup P$ is
equivalent to $P_2 \cup P$ for every program $P$.
\end{definition}

If two programs are strongly equivalent, we know that one of them can be replaced
with the other in a larger program without changing the declarative semantics. It
is clear that strong equivalence implies equivalence, but the converse
is not true. 

\begin{example}
Consider the programs $P_1= \set{a\lif \lnot b}$ and $P_2= \set{a}$, they are 
equivalent because $\set{a}$ is the unique answer set for both programs. However
$P_1\cup\set{b\lif a}$ has no answer sets, while $P_2\cup\set{b\lif a}$ has the answer
set $\set{a,b}$.
\end{example}

As a result of the study of strong equivalence of logic programs, an important
relation between the answer set semantics and intermediate logics appeared in the
following theorems.

\begin{theorem}\cite{LiPeVa01}\label{strongstable}
Let $P_1$ and $P_2$ be two augmented programs. Then $P_1$ and $P_2$ are strongly
equivalent iff $P_1$ and $P_2$ are equivalent in $\Gi[3]$ logic.
\end{theorem}

One intended use of this equivalence definition is to simplify programs.
We can, for instance, translate an augmented program into a free program preserving
strong equivalence. 

\begin{definition}\label{DefAugFree}
Let $P$ be an augmented program. Using distributive properties of conjunction,
disjunction and negation (all of them valid in $G_3$ logic) rewrite each clause in $P$
in the form $H \lif B$, where $H$ is a conjunction of simple disjunctions and $B$
is a disjunction of simple conjunctions.

Using the following equivalences we can eliminate conjunctions in the head of 
clauses, disjunctions in the body and atoms with two (or more) negations:
$$\begin{array}{l@{{}\equivin{G_3}{}}l}
A \land B \lif C & \begin{array}{l} A \lif C \\ B \lif C \end{array} \\[10pt]
A \lor \neg\neg B \lif C & A \lif \neg B \land C 
\end{array}
\qquad
\begin{array}{l@{{}\equivin{G_3}{}}l}
A \lif B \lor C  & \begin{array}{l} A \lif B \\ A \lif C \end{array} \\[10pt]
A \lif \neg\neg B \land C & A \lor \neg B \lif C
\end{array}
$$
We write $\AugFree{P}$ to denote the resulting free program.
\end{definition}

\begin{example}
We present now an example to explain how to compute the program
$\AugFree{P}$ for a given program $P$.
Suppose that we have the following augmented program $P$:
\begin{lpblock}
\lnot (a \land \lnot b) \land c \lif d \land (e \lor \lnot f). \\
\end{lpblock}
We can introduce negations into subformulas, applying
distributive properties of negation, until negation only appears in front
of atoms:
\begin{lpblock}
(\lnot a \lor \lnot \lnot b) \land c \lif d \land (e \lor \lnot f). \\
\end{lpblock}
Now, using distributive properties of conjunction and disjunction, we can write
the head (resp.\ body) of clauses in their normal conjunctive
(resp.\ disjunctive) form:
\begin{lpblock}
(\lnot a \lor \lnot \lnot b) \land c \lif (d \land e) \lor (d \land \lnot f). \\
\end{lpblock}
The head consists now of a conjunction of disjunctions. Using one of the
proposed equivalences we can remove all this conjunctions:
\begin{lpblock}
\lnot a \lor \lnot \lnot b \lif (d \land e) \lor (d \land \lnot f). \\
c \lif (d \land e) \lor (d \land \lnot f). \\
\end{lpblock}
Similarly, we proceed to remove disjunctions in the body:
\begin{lpblock}
\lnot a \lor \lnot \lnot b \lif d \land e. \\
\lnot a \lor \lnot \lnot b \lif d \land \lnot f. \\
c \lif d \land e. \\
c \lif d \land \lnot f. \\
\end{lpblock}
We can finally remove atoms with two (or more) negations using
the proposed equivalences:
\begin{lpblock}
\lnot a \lif \lnot b \land d \land e. \\
\lnot a \lif \lnot b \land d \land \lnot f. \\
c \lif d \land e. \\
c \lif d \land \lnot f. \\
\end{lpblock}
This program obtained corresponds to what we call $\AugFree{P}$.
\end{example}

An immediate consequence, obtained by the construction of $\AugFree{P}$, is an 
equivalence relation with respect to the logic $\Gi[3]$.

\begin{proposition}\cite{OsNaAr:lopstr}\label{augmentado-free}
Let $P$ be an augmented program. 
Then $P$ is equivalent under $\Gi[3]$ logic to the free program $\AugFree{P}$.
\end{proposition}

Using the machinery of logic we can conclude, 
from Theorem~\ref{strongstable} and Proposition~\ref{augmentado-free} above, that the 
defined transformation preserves strong equivalence. Formally we state the following 
theorem.

\begin{theorem}\cite{LiTaTu99}\label{AugToFree}
Let $P$ be an augmented program. Then $P$ is strongly equivalent to the 
free program $\AugFree{P}$.
\end{theorem}

\citeN{LiTaTu99} showed, using a similar transformation, that augmented programs can be
translated into free programs without changing the corresponding answer sets. Just
observe that the authors in \citeN{LiTaTu99} use the term ``equivalence'' to denote
a ``strong equivalence'' as we introduced it here. We emphasize the fact that this
result, with the language restricted to one kind of negation, can be obtained very easily
through equivalence relations in logic.

\begin{example}\label{aug2freesample}
Consider the following augmented program $P$:
\begin{lpblock}
a \lif \lnot\lnot a. \\
\lnot b \lif c \lor b.
\end{lpblock}
It is possible to construct, applying the rules described in Definition~\ref{DefAugFree},
a free program which, by Theorem~\ref{AugToFree}, is strongly equivalent to $P$. The
program $\AugFree{P}$ obtained is:
\begin{lpblock}
a \lor \lnot a. \\
\lnot b \lif c. \\
\lnot b \lif b.
\end{lpblock}
\end{example}

\subsection{Conservative Extensions}
If we want to use a program $P_2$ instead of another one $P_1$ it will be 
perfect if both programs have the same answer sets, but this condition is 
sometimes too much to expect. It will suffice, however, if we can identify 
through a simple relation the answer sets of the first program knowing those of
the second. A \emph{conservative extension} \cite{OsNaAr:lopstr} is one form of 
this weaker type of equivalence. 

\begin{definition}
Given two programs $P_1$ and $P_2$, we say that $P_2$ is a \emph{conservative extension} 
of $P_1$ if it holds that $M_1$ is an answer set of $P_1$ iff $M_2$ is an answer set of 
$P_2$, where $M_1$ and $M_2$ satisfy $M_1 = M_2 \cap \sig_{P_1}$.
\end{definition}

Note that our definition is different from that in \citeN{Bar03}, 
since we do not ask for  $P_1 \subseteq P_2$ to hold.
In order to define a ``strong'' version of this equivalence notion we find it useful to 
split the signature of atoms, used to construct logic programs, into two disjoint sets 
$\sig_U$ and $\sig_R$ that we call the \emph{user} and \emph{reserved} signature 
respectively. Unless stated otherwise, we assume that logic programs are restricted to 
the user signature, such programs are called \emph{user} programs. A program that is 
allowed to contain reserved atoms is called an \emph{internal} program.

\begin{definition}
Given a user program $P_1$ and an internal program $P_2$, we say that $P_2$ is a 
\emph{strong conservative extension} of $P_1$ if for every user program $P$, it holds 
that $P_2 \cup P $ is a conservative extension of $P_1 \cup P$.
\end{definition}

The idea is that users are allowed to write programs using only atoms from the user 
signature. The reserved signature will be used when new atoms are needed to perform 
internal program transformations in order to, for instance, simplify the structure of 
programs and compute answer sets. The notion of strong conservative extension allows to 
apply such transformations locally to fragments of programs.

A well-known transformation, that preserves this kind of equivalence, has been used to 
translate general programs into disjunctive ones.

\begin{definition}
Given a general program $P = D \cup C$, written as a disjoint union where
 $D$ is a disjunctive program and $C$ the set of constraints in $P$. We define 
 $\GenDisj{P} = D \cup \set{p \lif B \land \lnot p \tq (\bot \lif B) \in C}$, 
 where $p$ is a new atom in $\sig_R$.
\end{definition}

The following lemma is a direct consequence of the behavior of this 
transformation, see \citeN{Bar03}.

\begin{lemma}\cite{Bar03}\label{GenToDisj}
Let $P$ be a general program. $\GenDisj{P}$ is a strong conservative extension of $P$.
\end{lemma}

\citeN{sakino98} showed that every free program can be 
transformed, through a conservative extension, into a general one. 
We use instead the more economical transformation presented in \citeN{OsNaAr:lopstr}.
Essentially, the same idea is presented in Definition 2 from \citeN{Jan01}.

\begin{definition}\cite{OsNaAr:lopstr}\label{FreeToGenDef}
Given a free program $P$, let $S$ be the set containing all atoms $a$ such that $\lnot a$ 
appears in the head of some clause in $P$, and let $\varphi$ be an injective function, 
$\varphi \colon S \to \sig_R$, that assigns a new reserved atom to each element in $S$. Let
$P'$ be the program obtained from $P$ by replacing each occurrence of $\lnot a$ with
$\varphi(a)$ for every atom $a \in S$, and let $\Delta_S = \bigcup_{a \in S} 
\set{\varphi(a) \lif \lnot a, \bot \lif a \land \varphi(a)}$. Then we define
$\FreeGen{P} = P' \cup \Delta_S$.
\end{definition}

Again, the following proposition is obtained as a direct consequence of 
results presented in \citeN{OsNaAr:lopstr}.

\begin{proposition}\cite{OsNaAr:lopstr}\label{FreeToGen}
Let $P$ be a free program. $\FreeGen{P}$ is a strong conservative extension of $P$.
\end{proposition}

Note that in this case, if we have already  determined answer sets of $P_2$, it is 
possible to easily recover answer sets for $P_1$ just by taking the set intersection of 
each model with $\sig_{P_1}$. It turns out that, in fact, if $M$ is an answer set of 
$P_1$ then $M_S = M \cup \varphi(S \setminus M)$ is an answer set of $P_2$.

\begin{example}\label{ejemplito1}
Let $P$ be the free program:
\begin{lpblock}
a \lor \lnot a.
\end{lpblock}
$\FreeGen{P}$ is the program: 
\begin{lpblock}
a \lor x. \\
x \lif \lnot a. \\
\bot \lif x \land a.
\end{lpblock}
Recall that $P$ has two answer sets $M_1 = \set{}$ and $M_2 = \set{a}$. We obtain, as 
expected, that $\FreeGen{P}$ has also two answer sets: $\set{x}$ and $\set{a}$.
\end{example}

Observe that if $P_2$ is obtained from $P_1$ by a finite sequence of strong conservative 
and/or strong equivalence transformations, then $P_2$ is also a strong conservative 
extension of $P_1$.

Using this transformations we can, starting from an augmented program $P$, construct $P_1 
= \AugFree{P}$, $P_2 = \FreeGen{P_1}$ and $P_3 = \GenDisj{P_2}$. This chain of 
equivalences show that augmented programs are not more expressive than disjunctive ones 
under the answer set semantics. This means that, if we are able to compute answer sets of 
simple disjunctive programs, we can easily compute answer sets of more elaborated 
programs up to the augmented type. Formally we state the following theorem.

\begin{theorem}\label{AugEqDis}
For every augmented program $P$ there is a disjunctive program $P'$ such that $P'$ is a
strong conservative extension of $P$.
\end{theorem}

This result has also been presented in \citeN{PeSaSc02} where a polynomial transformation,
based on a technique that involves renaming subformulas, is used instead of our $\AugFree{P}$.
They even presented a working 
implementation\footnote{\url{http://www.cs.uni-potsdam.de/~torsten/nlp/}}
and proved nice properties like modularity
(a consequence of the transformation being a strong conservative extension). For theoretical
purposes any of these two transformations is equally valid in the following discussions.

\section{Characterization of Answer Sets}\label{sect-appl}

In this section we present one of our main results in Theorem~\ref{mainR},
which provides a characterization of answer sets of augmented programs in
terms of intuitionistic logic and we propose a definition
of answer sets for general propositional theories.

Given a signature $\sig$ and a set of atoms 
$M \subseteq \sig$ we define the complement of $M$ as $\setcomp{M} = \sig \setminus M$. 
The set $\sig$ is not always given explicitly, we assume $\sig = \sig_P$ when a program 
$P$ is clear by context.

Pearce provided a first characterization of answer sets in terms of intuitionistic logic.
He proved, at his Theorem~3.4 in~\citeyear{Pea99}, that a formula is entailed by a 
\emph{dis\-junc\-ti\-ve} program in the answer set semantics if and only if it 
belongs to every intuitionistically complete and consistent extension of the 
program formed by adding only negated atoms.

\begin{theorem}\cite{Pea99}\label{Pearce}
Let $P$ be a disjunctive program. (i) If $M$ is an answer set of $P$, then
$P \cup \lnot(\sig_P \setminus M)$ is intuitionistically consistent and complete.
(ii) Let $P \cup \Delta$ be intuitionistically consistent and complete, where
$\Delta \subseteq \lnot\sig_P$; then $\set{a \in \sig_P \tq P \cup \Delta \provein{I} a}$
is an answer set of $P$.
\end{theorem}

Using our notation this theorem states that a set of atoms $M$ is an answer set of $P$
if and only if $P \cup \lnot\setcomp{M} \cprovein{I} M$. This same result also holds for
the class of general programs, we can allow the use of constraints. However, it fails to
characterize answer sets if we allow negation in the head of clauses (free programs). Take
for instance the free program $P = \set{a \lor \lnot a}$.
According to Definition~\ref{answersetdef} this program has two answer sets: $\set{a}$ and
$\emptyset$. But only $\emptyset$, which corresponds to
$\lnot \setcomp{M} = \set{\lnot a}$, satisfies Pearce's condition. For the
other case the condition is reduced to $a \lor \lnot a \provein{I} a$, but this is not even
possible in classical logic.

We will see in the next section that the original approach from Pearce is actually
characterizing another important notion in ASP, answer sets satisfying the condition
in Theorem~\ref{Pearce} are also minimal models.

However, to actually obtain the answer sets of a program, according to
Definition~\ref{answersetdef}, we propose to extend it not only with negated atoms,
but allow twice negated atoms too. In previous example we would have
$a \lor \lnot a, \lnot \lnot a \provein{I} a$, recovering the answer set $\set{a}$.
We prove that this idea actually characterizes the notion of answer sets up to the
class of augmented programs.

\begin{theorem}\label{mainR}
Let $P$ be an augmented program and $M$ be a set of atoms. $M$ is an answer set of $P$ if 
and only if $P \cup \lnot\setcomp{M} \cup \lnot\lnot M \cprovein{I} M$.
\end{theorem}

This enhanced version of the theorem characterizes the notion of 
answer sets for augmented programs, and also provides a natural way to extend the 
definition of answer sets for logic programs containing arbitrary propositional formulas. 
Recall that the current definition of answer sets can only be applied to augmented
programs, while the intuitionistic statement in Theorem~\ref{mainR} does not seem
to imply any particular condition on the syntax of formulas in the program $P$.
This will allow, for instance, the use of embedded implications inside clauses
that were not allowed in augmented programs.

Here we will sketch the idea of the proof followed by an example constructed over a 
particular instance. The main idea is to reduce augmented programs, using transformations 
described in the previous section, into disjunctive programs where we use 
Pearce's result as a starting point.

Suppose we have an \emph{augmented} program $P$. We obtain first a \emph{free} program 
$P_1 = \AugFree{P}$ by unwinding clauses in $P$. Now, negation in the head of clauses in 
$P_1$ can be eliminated to obtain a \emph{general} program $P_2 = \FreeGen{P_1}$. Finally, 
constraints are removed to finish with a purely \emph{disjunctive} program $P_3 = 
\GenDisj{P_2}$.

As a consequence of equivalence theorems of previous section, answer sets of 
our disjunctive program $P_3$ are related, by a simple one to one relation, 
with answer sets of $P$. We can apply the result from Pearce, 
Theorem~\ref{Pearce}, to the disjunctive program $P_3$ and traverse the chain 
of transformations backwards in order to recover the original program $P$.

First we observe that answer sets of $P_2$ satisfy the same condition given 
by Pearce. This fact is obtained applying the following lemma to $P_2$.

\begin{lemma}\label{DisjToGen}
Let $P$ be a general program and $M$ be a set of atoms. \\ 
$\GenDisj{P} \cup \lnot (\sig_{\GenDisj{P}} \setminus M) \cprovein{I} M$ 
if and only if
$P \cup \lnot(\sig_P \setminus M ) \cprovein{I} M$.
\end{lemma}

Now, for the class of general programs, we can prove that both characterizations 
--proposed in Theorems~\ref{Pearce} and~\ref{mainR}-- coincide. Formally we state the 
following lemma.

\begin{lemma}\label{AddNN}
Let $P$ be a general program and $M$ be a set of atoms. \\
$P \cup \lnot\setcomp{M} \cprovein{I} M$ if and only if
$P \cup \lnot\setcomp{M} \cup \lnot\lnot M \cprovein{I} M$.
\end{lemma}

The crucial step in the proof is the following lemma, it allows us to remove additional 
atoms added to the language of $P_2$ when using the transformation $\FreeGen{P_1}$. This
would not be possible if we do not include the set $\lnot \lnot M$ to extend
the program. The set $S$ is obtained as in Definition~\ref{FreeToGenDef}, also recall
that (by Proposition~\ref{FreeToGen}) the answer sets of $P$ and $\FreeGen{P}$ are 
related by the identity $M_S = M \cup \varphi(S \setminus M)$.

\begin{lemma}\label{GenToFree}
Let $P$ be a free program and $M$ be a set of atoms. \\
$\FreeGen{P} \cup \lnot (\sig_{\FreeGen{P}} \setminus M_S) \cup \lnot\lnot M_S \cprovein{I} M_S$ if and only if \\
$P \cup \lnot (\sig_P \setminus M)  \cup \lnot\lnot M \cprovein{I} M$.
\end{lemma}

The final step is more simple, since the transformation of augmented to free programs 
already has some nice properties in terms of the $\Gi[3]$ logic. For our characterization 
we only need to show:

\begin{lemma}\label{FreeToAug}
Let $P$ be an augmented program and $M$ be a set of atoms. \\
$\AugFree{P} \cup \lnot\setcomp{M} \cup \lnot\lnot M \cprovein{I} M$
if and only if $P \cup \lnot\setcomp{M} \cup \lnot\lnot M \cprovein{I} M$.
\end{lemma}

Note that the language of $\AugFree{P}$ and $P$ is the same.
Following the chain of implications we are able to state that $M$ is an answer set of the 
original $P$ if and only if $P \cup \lnot\setcomp{M} \cup \lnot\lnot M \cprovein{I} M$. 
That is our Theorem~\ref{mainR}. We clarify the idea of the proof with a concrete 
example.

\begin{example}
Consider again the augmented program $P:$
\begin{lpblock}
a \lif \lnot\lnot a. \\
\lnot b \lif c \lor b.
\end{lpblock}

As we know from Example~\ref{ansetsample} the set $M = \set{a}$ is an answer set
for this program. Following Theorem~\ref{AugToFree}, as done in
Example~\ref{aug2freesample}, we construct the equivalent free program
$P_1 = \AugFree{P}$:
\begin{lpblock}
a \lor \lnot a. \\
\lnot b \lif c. \\ 
\lnot b \lif b.
\end{lpblock}

For this program we will replace atoms in $S = \set{a, b}$ that appear negated
in the head of clauses with new atoms, as in Proposition~\ref{FreeToGen}, 
to build a general program, which is still equivalent. This program $P_2 = \FreeGen{P_1}$ will 
contain
\begin{lpblock}[l@{\qquad\qquad}l]
a \lor x. & x \lif \lnot a. \\
y \lif c. & \bot \lif a \land x. \\
y \lif b. & y \lif \lnot b. \\
          & \bot \lif b \land y.
\end{lpblock}
with $M_S = \set{a, y}$ as the corresponding answer set. The final transformation 
$P_3 = \GenDisj{P_2}$ leads to the fully disjunctive program:
\begin{lpblock}[l@{\qquad\qquad}l]
a \lor x. & x \lif \lnot a. \\
y \lif c. & p \lif a \land x \land \lnot p. \\
y \lif b. & y \lif \lnot b. \\
          & p \lif b \land y \land \lnot p.
\end{lpblock}

Now we can apply Theorem~\ref{Pearce} from Pearce and obtain a proof 
for the intuitionistic claim $P_3 \cup \set{\lnot b, \lnot c, \lnot x}\cprovein{I} 
\set{a,y}$.
First, we can apply Lemma~\ref{DisjToGen} to obtain $P_2 \cup 
\set{\lnot b, \lnot c, \lnot x}\cprovein{I} \set{a,y}$. Now, according to 
Lemma~\ref{AddNN}, we can include the facts $\lnot\lnot M_S$ in the intuitionistic 
formula
$P_2 \cup \set{\lnot b, \lnot c, \lnot x, \lnot\lnot a, \lnot\lnot y} \cprovein{I} 
\set{a,y}$.

Recall that $P_2 = \FreeGen{P_1} = P'_1 \cup \Delta_S$ as written above. We can replace 
$P'_1$ with $P_1$, as described in the proof of Lemma~\ref{GenToFree}, to obtain the 
proof
$$P_1 \cup \set{x \lif \lnot a, \bot \lif a \land x, y \lif \lnot b, \bot \lif b \land y, 
\lnot b, \lnot c, \lnot x, \lnot\lnot a, \lnot\lnot y} \cprovein{I} a\,.$$
The atoms $x$ and $y$ added to the language of program when doing the transformation can 
now be mapped to the symbols $\bot$ and $\top$ respectively. We use this trick to 
eliminate them from the proof and to obtain:
$$P_1 \cup \set{\bot \lif \lnot a, \bot \lif a \land \bot, \top \lif \lnot b, \bot \lif b 
\land \top, \lnot b, \lnot c, \lnot \bot, \lnot\lnot a, \lnot\lnot \top} \cprovein{I} 
a\,.$$
This substitution works since, after some reductions, clauses originally contained in 
$\Delta_S$ are shown to be equivalent either to theorems or premises already listed. For 
this particular example formulas reduce to:
$$P_1 \cup \set{\lnot\lnot a, \top, \top , \lnot b}
\cup \set{\lnot b, \lnot c, \top} \cup \set{\lnot\lnot a, \top} \provein{I} a.$$
After removing such theorems, duplicate premises, and replacing $P_1$ with the original 
$P$, by Lemma~\ref{FreeToAug}, we finally obtain $P \cup \set{\lnot b, \lnot c} \cup 
\set{\lnot\lnot a} \cprovein{I} a$.
\end{example}

The characterization provided by Theorem~\ref{mainR} has several important consequences. 
We have as an immediate result a characterization of equivalence of logic 
programs (under the answer set semantics) in terms of intuitionistic logic.

\begin{corollary}\label{last}
Let $P_1$ and $P_2$ be two augmented programs sharing the same signature $\sig$. $P_1$ 
and $P_2$ are equivalent if and only if, for every set of atoms $M\subseteq \sig$,
$P_1\cup\lnot\setcomp{M}\cup\lnot\lnot M \equivin{I}P_2\cup\lnot\setcomp{M}\cup\lnot\lnot 
M$.
\end{corollary}

Another nice feature is that this characterization allows us to generalize the 
notion of answer sets to programs containing arbitrary propositional formulas 
as clauses. We propose the following definition.

\begin{definition}\label{Iansets}
Let $P$ be a logic program and $M$ be a set of atoms. $M$ is an \emph{answer set} 
of $P$ if $P \cup \lnot\setcomp{M} \cup \lnot\lnot M \cprovein{I} M$.
\end{definition}

A similar extension for the notion of answer sets for arbitrary theories can be
found at~\citeN{LiPeVa01}. They propose a generalization of answer sets in terms of
the equilibrium logic introduced by~\citeN{Pea99:negation}. It is an interesting
question left open to determine whether this two approaches are equivalent when
dealing with arbitrary propositional theories. Our conjeture is that they, indeed,
coincide.

In Section~\ref{Future}, we will discuss in detail several benefits and consequences of
such a definition. The important thing to observe is that this proposed definition
provides a methodology of representing knowledge in a  uniform way in the very well known
intuitionistic logic, where known theoretical  results can be applied to produce new
interesting effects. Now, programs with implication in the body have a meaning (following
Definition~\ref{Iansets}) and we can explore their use. Michael Gelfond points out (e-mail
communication) that ``the ability to use implication in the body seems to suggest the
following translation:
\begin{lpblock}[l@{\qquad}l]
\mbox{$r$ is true if every element with property $p$ has property $q$.} & \mbox{(*)} \\
\mbox{(Assume that the universe is finite)} & \\
\end{lpblock}
The natural translation is
\begin{lpblock}
\forall X (p(X) \rightarrow q(X)) \rightarrow r. \\
\end{lpblock}
If no implication is allowed in the formal language the translation 
of this English statement it loses its universal character. 
It now depends on the context and is prone to error.''
Hence, we believe that the use of our language could help to solve practical 
problems of representing knowledge. 
This makes sense since statements of type
(*) are very frequent. In our language, a formula of the form
$\forall X \alpha(X)$ could be introduced as an abbreviation of the conjunctive
formula $\alpha(a_1) \land \dots \land\alpha(a_n)$, where $\{a_1,\dots, a_n\}$ is
the Herbrand Universe of the program.

\section{Answer Sets and Minimal Models}\label{sect-minimal}

In this section we consider two valued interpretations, models and minimal models as
usual in logic programming, see \citeN{Llo87}. Minimal models are of general interest
for, at least, the following reasons: First, every answer set of a general program is a
minimal model. Second, they are closely related to circumscription
\cite{GePrPr89,MinPer85} and default logic \cite{LobSub92}. Third, they are of
theoretical and practical interest for a large class of optimization problems
\cite{Lib99}. Finally, computation of minimal models can be the first step towards 
computing answer sets of general programs, see for instance \citeN{BeNeNg93}.

In a few words we can say that the set $M$ is a minimal model of $P$ if $M$ is a model of 
$P$ (with respect to classical logic), and it is minimal (with respect to set inclusion)
among all other models of $P$. First of all, we provide a characterization of this notion
in terms of provability in classical logic.

\begin{lemma}\label{pearceconsistencia}
For a given augmented program $P$, $P \cup\lnot\setcomp{M}\cprovein{C} M$ iff $M$ is a 
minimal model of $P$.
\end{lemma}

We say that a set of atoms is a min-answer set if it is simultaneously a minimal model and 
an answer set. The following is also a main result of the paper. It provides a 
characterization of min-answer sets in terms of intuitionistic logic.

\begin{theorem}\label{charact-min-stable}
Let $P$ be an augmented program. $M$ is a min-answer set of $P$ iff $P 
\cup\lnot\setcomp{M}\cprovein{I} M$.
\end{theorem}

Observe that this is the same condition given by Pearce in Theorem~\ref{Pearce}. We 
conclude that he was actually characterizing min-answer sets and not answer sets in 
general. It turns out that answer sets of disjunctive programs are always minimal models 
and thus both characterizations coincide on this restricted class of programs.

We also want to note that min-answer sets are not the same thing as minimal answer sets. 
By a minimal answer set we understand an answer set which is minimal among all answer 
sets of a program. The next proposition and the following example should make this 
difference clear.
 
\begin{proposition}\label{minandstable}
If $M$ is a min-answer set of $P$, then $M$ is a minimal answer set of $P$.
\end{proposition}

\begin{example}
The converse of Proposition~\ref{minandstable} is not true. Let $P$ the program:
\begin{lpblock}
a \lor \lnot a. \\
b \lif a. \\
b \lif \lnot b.
\end{lpblock}
The unique answer set of $P$ is $\set{a,b}$ and, since it is unique, it is also a minimal 
answer set of $P$. But $\set{a, b}$ is not a minimal model, since $\set{b}$ is the unique 
minimal model of $P$, hence $P$ has no min-answer sets.
\end{example}

As a corollary of Theorems~\ref{AugEqDis} and~\ref{charact-min-stable} we can conclude 
that the class of answer sets is not more expressive than the class of 
min-answer sets.

\begin{corollary}\label{minimaleqstable}
For every augmented program $P$, there exists a computable disjunctive
program $P'$ such that the min-answer sets of $P'$, restricted
to $\sig_P$, are each and every answer set of $P$.
\end{corollary}

The following theorem provides a characterization of strong equivalence, 
si\-mi\-lar to the one in Theorem~\ref{strongstable}, for the class of min-answer sets.
We observe that the logic $\Gi[3]$ can be used, again, to test for strong equivalence.

\begin{theorem}\label{equi-str-min-stable}
Let $P_1$ and $P_2$ be two logic programs. Then $P_1$ and $P_2$ are strongly equivalent 
with respect to the min-answer set semantics if and only if $P_1$ and $P_2$ are 
equivalent in the logic $\Gi[3]$.
\end{theorem}

The proof is more complicated than the one needed to prove Theorem~\ref{strongstable}
in \citeN{LiPeVa01}, as we observe in Example~\ref{ex-seq}. Moreover, we use the 3-valued
logic $\Gi[3]$ instead of the Kripke semantics for HT.

\begin{example}\label{ex-seq}
Consider the programs $P_1 = \set{a \lif a}$ and $P_2 = \set{a \lor \lnot a}$.
These two programs are not strongly equivalent with respect to the answer set
semantics simply because they are not equivalent. For the min-answer set semantics the 
situation is more complicated. Both programs are equivalent because $M = \emptyset$ is 
the unique min-answer set of each program.
However, they are not strongly equivalent because, if we add the clause
$P = \set{a \lif \lnot a}$ to each program, the two programs are no longer equivalent
since $P_1 \cup P$ has no min-answer sets while $P_2 \cup P$ has exactly
one min-answer set: $\set{a}$.
\end{example}

It is easy to verify, using the characterization of min-answer sets in 
Theorem~\ref{charact-min-stable}, that programs equivalent under $\Gi[3]$ are strongly 
equivalent. For the converse we use the following two propositions that will allow us, 
under the assumption that two programs are not equivalent in $\Gi[3]$, to construct 
a third program that, when appended to the first two, will break equivalence with respect 
to the semantic of the min-answer sets.

\begin{remark}\label{G3-eq}
Let $A$ and $B$ be two formulas.
If $A \not\equivin{\Gi[3]} B$ then there is a 3-valued interpretation $I$ which
models $A$ and not $B$ (or models $B$ and not $A$).
\end{remark}

\begin{proposition}\label{main-G3}
Let $P_1$ and $P_2$ be arbitrary programs.
If there is a 3-valued interpretation $I$ such that $I$ models $P_1$ and does not
model $P_2$ then there exists a program $P$ such that: (i) $P_1 \cup P$ is
consistent and complete while $P_2 \cup P$ is inconsistent;
or (ii) $P_1 \cup P$ is incomplete and cannot be completed (preserving consistency) 
by adding only negated atoms while $P_2\cup P$ is both consistent and complete.
\end{proposition}

The following is an example to illustrate Proposition~\ref{main-G3}.

\begin{example}\label{ejemplito10}
Consider the programs $P_1 = \set{a \lif a}$ and $P_2 = \set{a \lor \lnot a}$.
Let $I$ be the 3-valued interpretation that evaluates $I(a)=1$. Observe that
$I$ models $P_1$ (since $I(P_1) = 2$) and does not model $P_2$ (since $I(P_2) = 1$).
Now if we take $P = \set{a \lif \lnot a}$ the program $P_1 \cup P$ becomes
incomplete (and unable to be completed by adding negated atoms) while $P_2\cup P$ is 
consistent and complete since it proves $a$. It turns out that $P_1 \cup P$ has 
no min-answer sets, while $P_2\cup P$ has the min-answer set $\set{a}$.
\end{example}
  
We end this section with a question that we have not been able to answer yet.
Suppose we have an augmented program $P$. Using Theorem 2 in~\citeN{PeSaSc02},
which states a transformation of augmented to disjunctive programs computable in
polynomial time that preserves the answer set semantics, we can compute the 
corresponding disjunctive program $P'$. 

Recall now that, in the restricted class of disjunctive programs, the semantics of
answer sets and min-answer sets coincide. Thus, under the assumption that
we have a min-answer set solver, we could compute the min-answer sets of $P'$ 
which are exactly the answer sets of $P'$. By the properties of the 
transformation in \citeN{PeSaSc02} we can recover, by a very simple 
transformation, the answer sets of the original program $P$.

Our question is if there is a similar, polynomial time computable, transformation
that could be used to compute min-answer sets for augmented programs under the assumption
that we have an answer set solver. In other words: is the class of min-answer sets more
expressive than the class of answer sets?
If the answer is no, which is our conjecture, then both semantics would be equivalent
in their power of representing problems. And there would be a great chance to obtain
feedback between these two paradigms. On the other hand, if the answer is yes, then
the min-answer sets would be more powerful than just answer sets. This would open a
new line of research on the class of problems that could be expressed using the
min-answer set semantics.

\section{Applications and Consequences}\label{Future}

There are several nice consequences of Theorem~\ref{mainR}. The first feature is 
that it provides a natural extension of the definition of answer sets for logic 
programs without depending on their particular restrictions of syntax or structure.
It has been a usual approach to restrict the language of logic programs to some
subsets of propositional logic while, the condition given in Theorem~\ref{mainR},
does not imply any of such restrictions. We could use now, for instance, embedded
implications in our programs, which are not allowed in the class of augmented
formulas.

The proposed Definition~\ref{Iansets} offers now an explanation of answer sets 
in terms of intuitionistic logic, where a wide variety of research has been done.
We explore here some ideas that, thanks to results shown in this paper, allow us
to better understand and generalize the notion of answer sets.

\subsection{Safe Beliefs}
Consider a logic agent, whose base knowledge of the world and its behavior is 
described by a set of propositional formulas $P$. Under the premise that $P$ is 
consistent, our agent can start infering from this base knowledge. 
Intuitionistic logic, a logic of knowledge, seems to be a natural inference system 
for this approach. We can say then that our agent \emph{knows} $F$, a 
propositional formula in general, if $P \cprovein{I} F$.

But we also want our agent to be able to do non-monotonic inference. Informally 
speaking we allow our agent to \emph{guess} or \emph{suppose} things in 
order to make more inference. But there is no reason, however, to just believe
everything that seems possible. We only \emph{suppose} facts if there is some 
reason to believe them or, more precisely, if they are helpful to produce any 
new knowledge.

Under this context we can rephrase the definition of answer sets. For this we 
introduce the symbol $\overline{M}$, the \emph{closure} of $M$, as 
$\overline{M} = M \cup \lnot \setcomp{M}$. This $\overline{M}$ contains a 
complete set of beliefs for our agent, the following definition states when 
this set of beliefs can be considered as safe.

\begin{definition}
Let $P$ be a logic program and $M$ be a set of atoms. Then $\overline{M}$ is a set 
of \emph{safe beliefs} if it satisfies $P \cup \lnot\lnot\overline{M} \cprovein{I} \overline{M}$.
\end{definition}

A very natural reading of the previous definition in the described context  is: 
\emph{``If a set of beliefs $\overline{M}$ is (i) consistent with the base knowledge 
and (ii) if we can suppose that the facts contained in $\overline{M}$ are true, and
this is enough to be sure about this facts, then it is safe to believe $\overline{M}$.''}
This \emph{suppose} corresponds to the double negation in the intuitionistic statement.
Observe that safe beliefs, defined this way, exactly correspond to the answer sets of $P$.

An immediate benefit of such a definition is that it extends the syntax of 
programs allowing embedded implication in clauses. This broader syntax can allow us
to write some rules for describing problems in a more natural way. We even 
suspect that this kind of syntax can be helpful to model concepts like 
aggregation in logic programs, as the ones described in \citeN{OsoJay99} and
\citeN{OsoJayPla99}. Further research has to be done in this direction in order to
present more concrete results.

On the theoretical point of view, there are also several benefits provided by 
this approach. Equivalence notions can be easily described in terms of logic. 
The fact that logic $\Gi[3]$ characterizes strong equivalence can be proved 
to hold for answer sets under this new definition. This is done in 
\citeN{Nav02} where a proof, that does not depend on the syntax of formulas, is given.

Other interesting result, presented in \citeN{OsNaAr:wollic}, is that the proposed 
definition of answer sets does not strictly depend on the underlying logic. 
It is proved that any proper intermediate logic does define the same semantic.

We try to demonstrate with this arguments various interesting possibilities on 
answer sets using logic. We show in \citeN{OsNaAr:iclp} how $\Gi[3]$ can be used
to debug a progam by taking advantage of the 3-valued nature of $\Gi[3]$.

\subsection{Answer Sets in other logics}

In this paper we do not consider the so-called classical negation, but it can easily
be included in the same intuitionistic framework by a simple renaming
method \cite{Bar03,GelLif90}. However, if we are interested in using this
classical negation with all of its power we can just replace intuitionistic
logic by a Nelson logic in our proposed Definition~\ref{Iansets},
see \citeN{OsPaAr02}.

Another interesting extension we can provide, due to this intuitionistic 
characterization, is a definition of answer sets for logic programs containing 
modal formulas. Modal logics were originated when trying to formalize notions 
like \emph{necessary} and \emph{possible} in logic. The new pair of connectives
$\lbox$ and $\ldiamond$ introduced have been also interpreted to model similar 
notions like tense, moral obligation and knowledge.

Using the well-known G\"odel embedding of intuitionistic logic into modal logic 
S4 we can provide a natural definition of answer sets based on this logic. 
The actual definition of the G\"odel mapping ${}^\circ$, that satisfies 
$\provein{I} A$ if and only if $\provein{S4} A^\circ$, can be found in
\citeN{ZaWoCh01}. 

We have to define some sort of \emph{basic acceptable knowledge} as unary formulas,
containing just one atom and unary connectives, that will play the role of the 
$\overline{M}$ above. For a modal logic program $P$ and a \emph{complete} set of
basic acceptable knowledge $M$ it would be reasonable to define $M$ as an 
\emph{answer set}, or \emph{safe beliefs} so to say, of $P$ if it satisfies 
$P \cup \lbox\ldiamond M \cprovein{S4} M$. More precise definitions have to be given,
but the main idea should be clear.

The use of modal formulas seems very appropriate since we would be able to 
explicitly model the concept of knowledge in logic programs. Then the generalization 
to multimodal logics should be a natural extension. This could provide a 
general framework where multiple agents can simultaneously reason, with the 
power of non-monotonic inference, about the knowledge and beliefs of each other.
Some advances in this research line are presented in \citeN{OsNaAr:aaai}.

A similar exercise can be done to define answer sets using linear logic. 
It is known that intuitionistic logic can be embedded in the propositional 
fragment of linear logic \cite{Gir87}. Thanks to our results and the given
embedding, it is possible to define the notion of an answer set in an 
environment with limited resources and thus provide some foundations for a 
framework of ASP as provability in linear logic, see \citeN{OsPaAr02}.

\section{Conclusions and Related work}\label{conclusion}

Work that relates ASP with classical logic can be found in
\citeN{EshKow89} and \citeN{NiePat96}. Erdem and Lifschitz relate answer sets
and supported models in \citeN{ErdLif01}. Work that relates ASP with
epistemic and modal logic can be found in \citeN{LifSch93} and \citeN{MarTru93}.
Our work follows the approach started by \citeN{Pea99}. We 
generalize his characterization, given for disjunctive programs, to
augmented programs. We also study the class of answer sets which are
also minimal models (min-answer sets) not done before to our knowledge.

We provide a characterization of answer sets in intuitionistic logic as
fo\-llows: a formula is entailed by an augmented program in the answer set
semantics if and only if it is proved in  every intuitionistically complete and
consistent extension of the program formed by adding only negated literals.
As we explain in the introduction, our result provides the foundations to
explain the notion of non-monotonic inference of any 
theo\-ry (using the standard connectives $\set{\lnot, \land, \lor, \lif}$) in 
terms of a monotonic logic (namely intuitionistic logic).

An immediate application of our result is to be able to have a definition
of ASP for arbitrary propositional theories. A similar result was presented
in~\citeN{LiPeVa01} where a generalization of answer sets in terms of the, so
called, equilibrium logic is stated. Our result provides, in particular, a 
natural way to extend the notion of answer sets in other logics.

Of particular interest to us are multimodal logics because they can be used to naturally
model the interaction of several agents. However, this is an open problem as we have
explained. We believe that, in general, our results presented in this paper re-emphasize
that the approach of answer sets is a solid paradigm to model non-monotonic 
reasoning.

We find a characterization of min-answer sets in terms
of intuitionistic logic. We observe that, in some way, the class
of answer sets is no more expressive than the class of min-answer sets.
We may ask: ``Is the class of min-answer sets more expressive than
the class of answer sets?'' We argue that, no matter what the answer is, it will
have impact in the theory of ASP.

Our results are given for propositional theories but they can easily be
ge\-ne\-ra\-li\-zed to universally quantified theories without functional
symbols. It would be interesting, however, to generalize our results to 
arbitrary first order theories.
Due to the large amount of knowledge in intuitionistic logic, we
expect to obtain a lot of feedback between these two areas.

\subsection*{Acknowledgements}
Numerous discussions with Michael Gelfond helped us to clarify our ideas.
This research is sponsored by the Mexican National Coun-cil
of Science and Technology, CONACyT (projects 35804-A and 37837-A).

\appendix
\section{Appendix: Proofs}\label{appendix-proofs}

In this section we will present references to some basic results and the
proofs of our main theorems and propositions in this paper.

\subsection{Basic Results}

The following are some basic results and definitions that were proved
in other sources and will be important for the proofs of our new results.

\begin{lemma}\cite{Dal80}\label{replace}
Let $T$ be any theory, and let $F, G$ be a pair of equivalent
formulas (under any intermediate $\name{X}$). Any theory obtained from $T$ by
replacing some occurrences of $F$ by $G$ is equivalent to $T$ (under logic
$\name{X}$).
\end{lemma}

\begin{lemma}\cite{OsNaAr:lopstr}\label{lang}
Let $T_1$, $T_2$ be two theories and let $A$ be a formula such that $\sig_{T_1
\cup \set{A}} \cap \sig_{T_2} = \emptyset$. If $T_2$ is a set of negative
literals and $T_1 \cup T_2 \provein{I} A$ then $T_1 \provein{I} A$.
\end{lemma}

\begin{definition}\cite{OsNaAr:lopstr}
The set $\mathbf{P}$ of \emph{positive formulas} is the smallest set
containing all formulas without negation connectives ($\neg$).
The set $\mathbf{N}$ of \emph{two-negated formulas} is the smallest set
 $\mathbf{X}$ with the properties:
\begin{enumerate}
\item If $a$ is an atom then $(\neg\neg a) \in \mathbf{X}$.
\item If ${A} \in \mathbf{X}$ then $(\neg\neg{A}) \in \mathbf{X}$.
\item If ${A}, {B} \in \mathbf{X}$ then $({A}\land{B}) \in \mathbf{X}$.
\item If ${A} \in \mathbf{X}$ and ${B}$ is any formula then $({A}\lor{B}), 
({B}\lor{A}), ({A}\lif{B}) \in \mathbf{X}$.
\end{enumerate}
For a given set of formulas $\Gamma$, we define the \emph{positive subset} of $\Gamma$, 
denoted $\name{Pos}(\Gamma)$, as the set $\Gamma \cap \mathbf{P}$.
\end{definition}

\begin{proposition}\cite{OsNaAr:lopstr}\label{nonegs}
Let $\Gamma$ be a subset of $\mathbf{P}\cup\mathbf{N}$, and let ${A} \in 
\mathbf{P}$ be a positive formula. If $\Gamma \provein{I} {A}$ then 
$\name{Pos}(\Gamma) \provein{I} {A}$.
\end{proposition}

\subsection{Proofs about equivalence}

\begin{proof}[Proof of Theorem~\ref{AugEqDis}]
Let $P$ be an augmented program. By Theorem~\ref{AugToFree}, $P_1 = \AugFree{P}$ is strongly equivalent to $P$. Then, by Proposition~\ref{FreeToGen}, $P_2 = \FreeGen{P_1}$ is a conservative extension of $P_1$. Similarly, by Lemma~\ref{GenToDisj}, $P_3 = \GenDisj{P_2}$ is a conservative extension of $P_2$. Through the chain of equivalences we obtain that $P_3$ is a conservative extension of $P$.
\end{proof}

\subsection{Reductions for General Programs}

We present here some definitions and simple results of reductions, motivated by results
in \citeN{DiOsZe01}, for the class of general programs. They are helpful in the proof of
Theorem~\ref{mainR}, particularly at Lemma~\ref{AddNN}.

\begin{definition}[First Reduction]\cite{OsNaAr:wollic}
Let $P$ be a general program and $M$ be a set of atoms. We define the
\emph{first reduction} of $P$ with respect to $\lnot M$, denoted
$\RedOne{P;\lnot M}$, as the program obtained applying the following
transformation to each clause $H \lif B$ contained in $P$:
\begin{itemize}
\item Delete from $B$ all literals $\lnot a$ such that $\lnot a \in \lnot M$.
\item Delete from $H$ all literals $a$ such that $\lnot a \in \lnot M$.
\item Delete the clause if there is some literal $a$ in $B$ such that 
$\lnot a \in \lnot M$.
\end{itemize}
\end{definition}

\begin{lemma}\label{p-redu2}
Let $P$ be a general program and $M$ be a set of atoms. The second reduction satisfies
$P \cup \lnot\lnot M \equivin{I} \RedTwo{P;\lnot\lnot M} \cup \lnot\lnot M$.
\end{lemma}

\begin{definition}[Second Reduction]
Let $P$ be a general program and let $M$ be a set of atoms. We define the
\emph{second reduction} of $P$ with respect to $\lnot\lnot M$, denoted
$\RedTwo{P;\lnot\lnot M}$, as the program obtained applying the following
transformation to each clause $H \lif B$ contained in the program $P$:
\begin{itemize}
\item If $H = \bot$ then delete from $B$ all literals $a$ such that $\lnot\lnot a \in 
\lnot\lnot M$.
\item Delete the clause if there is some literal $\lnot a$ in $B$ such that $\lnot\lnot a 
\in \lnot\lnot M$.
\end{itemize}
\end{definition}

\begin{lemma}\cite{OsNaAr:wollic}\label{p-redu1}
Let $P$ be a general program and $M$ be a set of atoms such that $P \cup \lnot M$ is 
consistent. If $P' = \RedOne{P;\lnot M}$ then the following properties hold:
\begin{enumerate}
\item $P \cup \lnot M \equivin{I} P' \cup \lnot M$.
\item $\sig_{P'} \cap M = \emptyset$.
\end{enumerate}
\end{lemma}

\begin{proof}
The two transformation steps in the reduction can be justified since it is possible to show, using intuitionistic logic, $\lnot\lnot a \provein{I} (\bot \lif a \land B) \liff (\bot \lif B)$ and $\lnot\lnot a \provein{I} H \lif \lnot a \land B$ respectively.
\end{proof}

\subsection{Proofs about the characterization of answer sets}

\begin{proof}[Proof of Lemma~\ref{DisjToGen}]
In the following paragraph the set $\setcomp{M}$ always represents the set 
$\sig_P \setminus M$, that is the set complement of $M$ with respect to
the signature of the program $P$. Observe that, making this assumption,
the set $\sig_{\GenDisj{P}} \setminus M = (\sig_P \cup \set{p}) \setminus M
= \setcomp{M} \cup \set{p}$.

The transformation step that defines $\GenDisj{P}$ preserves equivalence in intuitionistic logic, since $\lnot p \provein{I} (\bot \lif B) \liff (p \lif B \land \lnot p)$.
So in particular $\GenDisj{P} \cup \lnot\setcomp{M} \cup \set{\lnot p} \cprovein{I} M$ iff $P \cup \lnot\setcomp{M} \cup \set{\lnot p} \cprovein{I} M$ iff, by Lemma~\ref{lang} and since $p \notin \sig_P$, $P \cup \lnot\setcomp{M} \cprovein{I} M$.
\end{proof}

\begin{proof}[Proof of Lemma~\ref{AddNN}]
First suppose $P\cup\lnot\setcomp{M} \cprovein{I} M$. Since $A \lthen \lnot\lnot A$ is 
an intuitionistic theorem $P\cup\lnot\setcomp{M} \cprovein{I} \lnot\lnot M$. Therefore
$P\cup\lnot\setcomp{M} \cup \lnot\lnot M$ is consistent and since intuitionistic logic
is monotone we have $P\cup\lnot\setcomp{M} \cup \lnot\lnot M \cprovein{I} M$.

\medskip
Now suppose $P\cup\lnot\setcomp{M}\cup\lnot\lnot M \cprovein{I} M$, it is
immediate that $P\cup\lnot\setcomp{M}$ is consistent.
Now, we want to show that $P\cup\lnot\setcomp{M} \provein{I} M$.
If $P' = \RedOne{P; \lnot\setcomp{M}}$ then, by Lemma~\ref{p-redu1}, $P \cup \lnot \setcomp{M} \equivin{I} P' \cup \lnot \setcomp{M}$ and $\sig_{P'} \cap \setcomp{M} = \emptyset$.

Since $P \cup \lnot\setcomp{M} \cup \lnot\lnot M \provein{I} M$, then $P' \cup \lnot\setcomp{M} \cup \lnot\lnot M \provein{I} M$. It is also clear that $\sig_{P' \cup \lnot\lnot M} \cap \sig_{\lnot\setcomp{M}} = \emptyset$, so we can apply Lemma~\ref{lang} to get $P' \cup \lnot\lnot M \provein{I} M$.

Let $P'' = \RedTwo{P'; \lnot\lnot M}$. Since $\sig_{P'} \cap \setcomp{M} = \emptyset$, that $\sig_{P'} \subset M$. So, by definition of ${\rm redu2}$, disjunctive clauses containing negative literals in $P'$ will be always removed. Also note that $P''$ can not contain constraints. If $P'$ contains a constraint of the form $\bot \lif B$ and is not removed is because only positive literals occur in $B$. However ${\rm redu2}$ will, for this clause, remove all literals in the body and leave a clause $\bot \lif \top$. But this is a contradiction, since we already know that $P \cup \lnot M \cup \lnot\lnot M$ is consistent.
So $P''$ is an entirely positive program and, moreover, $P'' \subset P'$.

Since, by Lemma~\ref{p-redu2}, $P' \cup \lnot\lnot M \equivin{I} P'' \cup \lnot\lnot M$ we have $P'' \cup \lnot\lnot M \provein{I} M$ and, using Proposition~\ref{nonegs}, $P'' \provein{I} M$.
But we already know that $P \cup \lnot\setcomp{M} \provein{I} P' \cup \lnot\setcomp{M}$ and, since $P'' \subset P'$, in particular $P \cup \lnot\setcomp{M} \provein{I} P''$. So finally we obtain, as desired, $P \cup \lnot\setcomp{M} \provein{I} M$.
\end{proof}

\begin{proof}[Proof of Lemma~\ref{GenToFree}]
Suppose $\FreeGen{P} \cup \lnot\setcomp{M_S} \cup \lnot\lnot M_S \cprovein{I} M_S$. Recall that $\FreeGen{P}$ can be written as $P' \cup \Delta_S$ and, since $P \cup \Delta_S \provein{I} P'$, $P \cup \Delta_S \cup \lnot\setcomp{M_S} \cup \lnot\lnot M_S \cprovein{I} M_S$. Since $M_S = M\cup\varphi(S\setminus M)$ we can break the sets $\lnot\setcomp{M_S}$ and $\lnot\lnot M_S$ as disjoint subsets from user and reserved atoms. That is $\lnot\setcomp{M_S} = \lnot\setcomp{M}\cup\lnot[\varphi(S\cap M)]$ and $\lnot\lnot M_S = \lnot\lnot M \cup \lnot\lnot[\varphi(S\setminus M)]$.\footnote{Note that $\setcomp{M} = \sig_P \setminus M$, while $\setcomp{M_S} = (\sig_{P} \cup \varphi(S)) \setminus M_S$}

Similarly we can write $\Delta_S$ as the disjoint union $\Delta_M \cup \Delta_{(S \setminus M)}$. Then, since $\lnot\lnot M \cup \lnot[\varphi(S\cap M)] \provein{I} \Delta_M$, we have that $$P\cup\Delta_{(S \setminus M)}\cup\lnot\setcomp{M}\cup\lnot\lnot M \cup \lnot[\varphi(S\cap M)] \cup \lnot\lnot[\varphi(S\setminus M)] \cprovein{I} M\,.$$

In the intuitionistic proof for each $a \in M$ as shown above we can map each
symbol $x \in \varphi(S\cap M)$ to $\bot$ and each symbol
$y \in\varphi(S\setminus M)$ to $\top$. This will lead to
a proof for $P\cup\lnot\setcomp{M}\cup\lnot\lnot M \provein{I} M$, since
premises in $\Delta_{(S \setminus M)}$, $\lnot[\varphi(S\cap M)]$ and
$\lnot\lnot[\varphi(S\setminus M)]$
are mapped either to intuitionistic theorems or elements in $\lnot\setcomp{M}$.

\medskip To prove the other implication assume $P \cup \lnot\setcomp{M} \cup \lnot\lnot M \cprovein{I} M$. From definition of $\Delta_S$ we have that $P \cup \Delta_S \cup \lnot\setcomp{M} \cup \lnot\lnot M$ is consistent, also note that $\Delta_S \cup \lnot\setcomp{M} \cup \lnot\lnot M \provein{I} \varphi(S\setminus M) \cup \lnot[\varphi(S\cap M)]$. It follows then that
$$P'\cup\Delta_S \cup \lnot\setcomp{M} \cup \lnot\lnot M \cup \lnot[\varphi(S\cap M)] \cup
\lnot\lnot[\varphi(S\setminus M)] \cprovein{I} M \cup \varphi(S\setminus M)$$
as we wanted.
\end{proof}

\begin{proof}[Proof of Lemma~\ref{FreeToAug}]
By construction we have $P \equivin{\Gi[3]} \AugFree{P}$. In particular, for every set of atoms $M$, this implies $P \cup \lnot\setcomp{M} \cup \lnot\lnot M \equivin{I} \AugFree{P} \cup \lnot\setcomp{M} \cup \lnot\lnot M$. So we finally obtain $P \cup \lnot\setcomp{M} \cup \lnot\lnot M \cprovein{I} M$ iff $\AugFree{P} \cup \lnot\setcomp{M} \cup \lnot\lnot M \cprovein{I} M$.
\end{proof}

\begin{proof}[Proof of Theorem~\ref{mainR}]
The set of atoms $M$ is an answer set of the augmented program $P$ iff, by Theorem~\ref{AugToFree}, $M$ is an answer set of $P_1 = \AugFree{P}$ iff, by Proposition~\ref{FreeToGen}, $M_S$ is an answer set of $P_2 = \FreeGen{P_1}$ iff, by
Lemma~\ref{GenToDisj}, $M_S$ is an answer set of $P_3 = \GenDisj{P_2}$ iff, by Theorem~\ref{Pearce}, $P_3 \cup \lnot \setcomp{M_S} \cprovein{I} M_S$ iff, by Lemma~\ref{DisjToGen}, $P_2 \cup \lnot \setcomp{M_S} \cprovein{I} M_S$ iff, by Lemma~\ref{AddNN}, $P_2 \cup \lnot \setcomp{M_S} \cup \lnot\lnot M_S \cprovein{I} M_S$ iff, by Lemma~\ref{GenToFree}, $P_1 \cup \lnot \setcomp{M} \cup \lnot\lnot M \cprovein{I} M$ iff, by Lemma~\ref{FreeToAug}, $P \cup \lnot \setcomp{M} \cup \lnot\lnot M \cprovein{I} M$.
\end{proof}

\begin{proof}[Proof of Corollary~\ref{last}]
We have that the two programs $P_1$ and $P_2$ are equivalent in the answer set semantics
iff, by definition, ($M$ is an answer set of $P_1$ iff $M$ is an answer set of $P_2$)
iff, by Theorem~\ref{mainR}, ($P_1 \cup\lnot\setcomp{M}\cup \lnot\lnot M\cprovein{I} M$
iff $P_2 \cup\lnot\setcomp{M}\cup \lnot\lnot M\cprovein{I} M$)
iff, since $P_1 \cup\lnot\setcomp{M}\cup \lnot\lnot M$ and
$P_2 \cup\lnot\setcomp{M}\cup \lnot\lnot M$ are literal complete theories,
we can conclude that $P_1\cup\lnot\setcomp{M}\cup\lnot\lnot M \equivin{I}
P_2\cup\lnot\setcomp{M}\cup\lnot\lnot M$.
\end{proof}

\subsection{Proofs about answer sets and minimal models}

\begin{proof}[Proof of Lemma~\ref{pearceconsistencia}]
Suppose that $P \cup\lnot\setcomp{M}\cprovein{C} M$. Then
$M$ is model of $P$. Suppose then that $M$ is not a minimal model of $P$.
Then there exists $N$, a model of $P$ such that $N \subset M$, take $a \in M\setminus N$.
In particular $\lnot a \in \lnot\setcomp{N}$, 
thus $P\cup\lnot\setcomp{N}\provein{I}\lnot a$. But, since
$P\cup\lnot\setcomp{M}\provein{I} a$ and $\lnot\setcomp{M} \subseteq \lnot\setcomp{N}$,
$P\cup\lnot\setcomp{N}\provein{I}\lnot a$ and $N$ is not a model.

For the converse: 
if $M$ is a minimal model of $P$ then $P$ is consistent (it has one model) 
and $P\cup\lnot\setcomp{M}$ is also consistent.
But it is easy to check that $M$, since it is minimal, is the unique
model of $P\cup\lnot\setcomp{M}$.
So $P\cup\lnot\setcomp{M}\provein{C} M$.
\end{proof}

\begin{proof}[Proof of Theorem~\ref{charact-min-stable}]
Suppose that $M$ is a min-answer set of $P$ so 
$P \cup\lnot\setcomp{M}\cup \lnot\lnot M\cprovein{I} M$, 
because $M$ is an answer set of $P$. Since $M$ is a minimal model,
we know that $P \cup\lnot\setcomp{M}\provein{C} M$, by
Lemma~\ref{pearceconsistencia}, then 
$P \cup\lnot\setcomp{M}\provein{I} \lnot\lnot M$. By the last assertion and since
$P \cup\lnot\setcomp{M}\cup \lnot\lnot M\cprovein{I} M$, we have
$P \cup\lnot\setcomp{M}\provein{I} M$ and $P \cup\lnot\setcomp{M}$
is consistent, i.e. $P \cup\lnot\setcomp{M}\cprovein{I} M$.
  
For the converse,
suppose $P \cup\lnot\setcomp{M}\cprovein{I} M$
so $P \cup\lnot\setcomp{M}\provein{I} \lnot\lnot M$,
hence $P \cup\lnot\setcomp{M}\cup \lnot\lnot M$ is consistent
in intuitionistic logic and $M$ is an answer set of $P$.
On the other hand, if $P \cup\lnot\setcomp{M}\cprovein{I} M$ 
then $P \cup\lnot\setcomp{M}\provein{C} M$ 
and we know that $P \cup\lnot\setcomp{M}$
is consistent in intuitionistic logic,
which implies consistency in classical logic.
By Lemma~\ref{pearceconsistencia}, 
we have that $M$ is a minimal model of $P$.
\end{proof}
 
\begin{proof}[Proof of Proposition~\ref{minandstable}]
Suppose that $M$ is a min-answer set of $P$, 
but $P$ is not a minimal answer set of $P$. Then
there is $N \subset M$,  
such that $N$ is a minimal answer set of $P$.
In particular $N$ is a model of $P$. 
But this is not correct, since $M$ is a minimal model of $P$.
\end{proof}

\begin{proof}[Proof of Corollary~\ref{minimaleqstable}]
Follows by Theorem~\ref{AugEqDis} and the well known fact that, for disjunctive
programs, the answer sets are minimal models.
\end{proof}

\begin{proof}[Proof of Proposition~\ref{main-G3}]
We assume that $I'$ is defined as in the proposition above. Observe in particular that, if $I$ models $A$ then $I'$ models $A$ too.

Consider also the following definition:
For a given interpretation $I$ in $\Gi[3]$ the program $T(I)$ is defined as the 
minimum set $X$ which satisfies
\begin{enumerate}[88.]
\item If $I(a) = I(b) = 1$ and $a \neq b$ then $(b\lif a) \in X$.
\item If $I(a) = 1$ then $(a \lif \lnot a) \in X$.
\item If $I(a) = 2$ then $(a) \in X$.
\item If $I(a) = 0$ then $(\bot \lif a) \in X$.
\end{enumerate}

We have two main cases in the proof of this proposition:

\medskip\noindent\textit{1.} There is a \emph{definite} interpretation $I$ that models
$P_1$ and not $P_2$. Let $P = T(I)$ (as just defined). Then, by construction,
$P_1 \cup P$ is a consistent and complete extension of $P_1$, while $P_2 \cup P$ is
inconsistent.

\medskip\noindent\textit{2.} If every interpretation that models $P_1$ and not $P_2$ is
indefinite, then let $I$ be one of such interpretations and $P = T(I)$.
Since $I$ models $P_1$ we have that $I'$ models $P_1$ too. 
Notice that $I \neq I'$ since $I$ contains some $1$ assignments and $I'$ not.

Since $I$ models $P$, we have $I$ models $P_1 \cup P$. 
Again $I'$ models $P_1 \cup P$ too. 
But $I \neq I'$ so there are two different interpretations that model
$P_1 \cup P$ and hence it is not complete. Nor it can be completed by adding 
negated atoms $\lnot a$ since the program will become inconsistent 
(if $I(a) = 1$ or $I(a) = 2$) or still be incomplete (if $I(a)=0$).

We prove now that $I'$ models $P_2$. If not $I'$ will model $P_1$ and not $P_2$ 
but $I'$ is definite contradicting the hypothesis of this case. 
Again, since $I'$ models $P$, $I'$ models $P_2 \cup P$.

\medskip It will be shown that $I'$ is the only interpretation that 
models $P_2 \cup P$. Suppose $K$ is another interpretation that models 
$P_2 \cup P$, and hence $K$ models $P_2$.

\textit{Case 1.} $I(a) = 2$. Then $a \in P \subset P_2 \cup P$. 
But $K(a) \neq 2$ will imply $K(P_2 \cup P) \neq 2$ contradicting the fact that 
$K$ models $P_2 \cup P$. So if $I(a) = 2$ then $K(a) = 2$.

\textit{Case 2.} $I(a) = 0$. Then $\lnot a \in P \subset P_2 \cup P$. 
But $K(a) \neq 0$ will imply $K(\lnot a) \neq 2$ and $K(P_2 \cup P) \neq 2$ 
contradicting the fact that $K$ models $P_2 \cup P$. So if $I(a) = 0$ then $K(a) = 0$.

\textit{Case 3a.} $I(a) = 1$ and $K(a) = 0$. Then 
$(\lnot a\lthen a) \in P \subset P_2 \cup P$. But $K$ will evaluate 
$K(\lnot a\lthen a) = 0$ and $K(P_2 \cup P) = 0$ arising contradiction again.

\textit{Case 3b.} $I(a) = 1$ and $K(a) = 1$. 
Then, if exists, take another atom $b$ such that $I(b) = 1$.
 Now $\set{a\lthen b, b \lthen a} \subset P \subset P_2 \cup P$ and, 
 since $K$ models $P_2 \cup P$, $K(a\liff b) = 2$, hence $K(a) = K(b) = 1$. 
 So in this case $I(a) = 1$ implies $K(b) = 1$ for all atoms $b$, leading to 
 $I = K$. But, from hypothesis, $I$ did not model $P_2$ and $K$ does. 
 Contradiction.

\textit{Case 3c.} Previous two cases state $I(a) = 1$ implies $K(a) = 2$ and, 
together with cases 1 and 2, are sufficient to imply $K = I'$. So, as claimed, $I'$ is
the only model for $P_2 \cup P$.
\end{proof}

\begin{proof}[Proof of Theorem~\ref{equi-str-min-stable}]
Let $P_1$ and $P_2$ be two logic programs.
If $P_1$ and $P_2$ are equivalent in $\Gi[3]$ then, for any $P$,
$P_1\cup P$ and $P_2\cup P$ are equivalent in $\Gi[3]$. 
We will prove assuming that $M$ is a min-answer set of $P_1 \cup P$ that it is also
a min-answer set of $P_2 \cup P$. Since $P_1 \equivin{\Gi[3]} P_2$ it is 
immediate that $M$ is an answer set of $P_2 \cup P$ by Theorem~\ref{strongstable}.

Now, since $M$ is a minimal model of $P_1 \cup P$, $P_1 \cup P \cup \lnot\setcomp{M} \cprovein{C} M$ but, in particular, $P_1 \cup P \equivin{C} P_2 \cup P$ and therefore $P_2 \cup P \cup \lnot\setcomp{M} \cprovein{C} M$. By Lemma~\ref{pearceconsistencia}, $M$ is a minimal model of $P_2 \cup P$.
The same argument proves that every min-answer set of $P_2 \cup P$ is a min-answer set of $P_1 \cup P$. So the two programs are strongly equivalent with respect to the
sematic of the min-answer sets.

\medskip
For the converse, suppose that $P_1$ and $P_2$ are not equivalent in $\Gi[3]$
then, by Remark~\ref{G3-eq}, there is an interpretation that models $P_1$, but 
not $P_2$ which, by Proposition~\ref{main-G3}, implies they are not strongly equivalent
with respect to the semantic of the min-answer sets.
\end{proof}

\bibliographystyle{acmtrans}
\bibliography{juan}

\end{document}